\documentclass[twocolumn,twoside,slac]{revtex4}
\usepackage{graphicx}
\usepackage{fancyhdr}
\begin{document}

\title{ An Unbinned Goodness-of-Fit Test Based on 
      \\ the Random Walk\footnote{University of Cincinnati preprint \# UCHEP-03-02}}


\author{K. Kinoshita}
\affiliation{University~of~Cincinnati \\
Cincinnati, OH 45221 USA}

\begin{abstract}
We describe a test statistic for unbinned goodness-of-fit of data in one dimension.  
The statistic is based on the two-dimensional Random Walk.
The rejection power of this test is explored both for simple and compound hypotheses and, for the examples explored, it is found to be comparable to that for the $\chi^2$ test.
We discuss briefly how it may be possible to extend this test to multi-dimensional data.

\end{abstract}

\maketitle

\thispagestyle{fancy}


\section{Introduction}
This search for an unbinned goodness-of-fit test has been motivated by the widespread use of unbinned maximum likelihood fitting for determining $CP$-violating parameters at  Belle.
While there are many cross-checks to insure that there are no spurious signals and biases, the fits tend to be complicated and not very transparent.
They often involve probability density functions (PDF's) that differ with every event, based on measured quantities that add dimensions to the data that are not explicit in the fits.
As there is no widely accepted unbinned goodness-of-fit test that applies to such fits,  testing for statistical consistency of results has been uneven.
The tests that have been done, resorting to binned $\chi^2$ or toy Monte Carlo, have their place but have not been entirely satisfactory in addressing the question.

A common technique of unbinned tests involves first transforming the measured quantities to a variable in which the null hypothesis has a uniform distribution, where the PDF is flat, and then to test this ``flattened'' distribution for consistency with uniformity.
%Flat data have the advantage that there is no question of scale invariance -- all statistical fluctuations from the null hypothesis are on equal footing.
There exists a variety of tests for uniformity, but most are not readily extended to multidimensional data,
and they do not address compound hypotheses.
A review of methods is given in \cite{aslandurham}.

In this report, 
we explore a test statistic that is based on the two-dimensional Random Walk.
To begin, its distribution in the case of a flat PDF is discussed.  
The ensemble distribution is then found for several alternate hypotheses, and the rejection power is calculated for comparison with other goodness-of-fit tests.
As the aim of a goodness-of-fit test as it would be applied at Belle is to test the validity of the parametrization used in fitting, it is also important to examine how the test is modified under compound hypotheses.  
The discussion is thus expanded to include data which are fitted to determine one or more parameters.
Finally, we discuss the possibility of extending to multidimensional data.

\section{Random Walk as a Test of Flatness}
A dataset consisting of $N$ measurements of the one-dimensional quantity $x$ lying in the interval $[0,1]$ may be mapped trivially to points on a unit circle with polar angle $\phi$ on the interval $[0,2\pi]$, so that each point is considered to be a unit vector with direction defined by $\phi$.
If the PDF in $x$ is flat, the vector sum of the corresponding unit vectors in two dimensions corresponds to the net displacement, $D$, after a two-dimensional Random Walk of $N$ steps with unit step size.
For sufficiently large $N$, this distribution converges to a well-known form (Rayleigh, 1888) and the distribution in $D^2$ is  an exponential decay with mean equal to $N$ .
We take $D^2/N$ as the test statistic.
A deviation of the root distribution from the hypothesis will result in a bias of the ensemble distribution of this test statistic away from the origin.
This statistic is mathematically equivalent to the first order term in the Fourier series that describes the distribution of the data:
\begin{eqnarray}
{\cal F}(k=1) &= &\int_0^{2\pi}d\phi \sum_{j=1}^N{e^{ik\phi}}\delta(\phi-\phi_j)\label{eq:fourier}\\
&=&\sum_{j=1}^N{e^{i\phi_j}}\nonumber
\end{eqnarray}
where one can see that $D^2\propto |{\cal F}(1)|^2$.
One would expect this distribution to be most sensitive to an overall imbalance of the PDF in generally opposite $\phi$ directions.  
To obtain sensitivity to higher order differences, one could thus take successively higher order terms in the series, for $k=2,...$.  
In practice it may not be useful to examine terms above $k=3$.
In this study we look at $k=1$  ($d=1$) and define ${K}_k\equiv \frac{|{\cal F}(k)|^2}{N}$.
What we have defined as $K_1$ appears in the review of D'Agostino and Stephens\cite{statistic} as $R$ in the context of the Von Mises test, a test for uniformity on a circle.

\section{Flat PDF}
As mentioned above, the ${K}_1$ distribution for a flat PDF converges rapidly to an exponential with a decay constant of unity.
Figure~\ref{fig:null} (top row) shows the distributions in ${K}_1$ for ensembles of randomly generated experiments containing $N=10$, 100, and 1000 events.
Each of the three distributions is fitted via binned maximum likelihood to an exponential form.
The fitted inverse decay constants (``slopes'') are $0.992\pm 0.010$, $1.008\pm 0.033$, and $1.039\pm 0.049$, respectively, in excellent agreement with the expectation.

\begin{widetext}
%%%%%%%%%%%%%%%%%%%%%%%%%%%%%%
\begin{center}
\begin{figure}[htb]
\smallskip
\includegraphics[width=135mm]{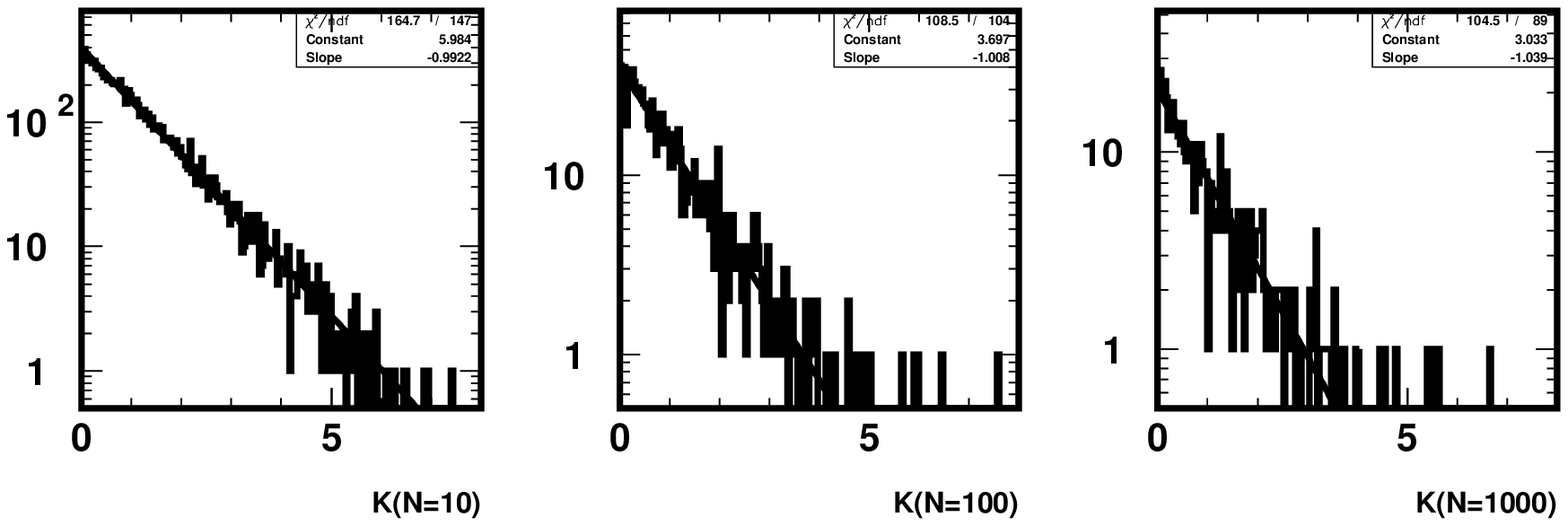}
\includegraphics[width=135mm]{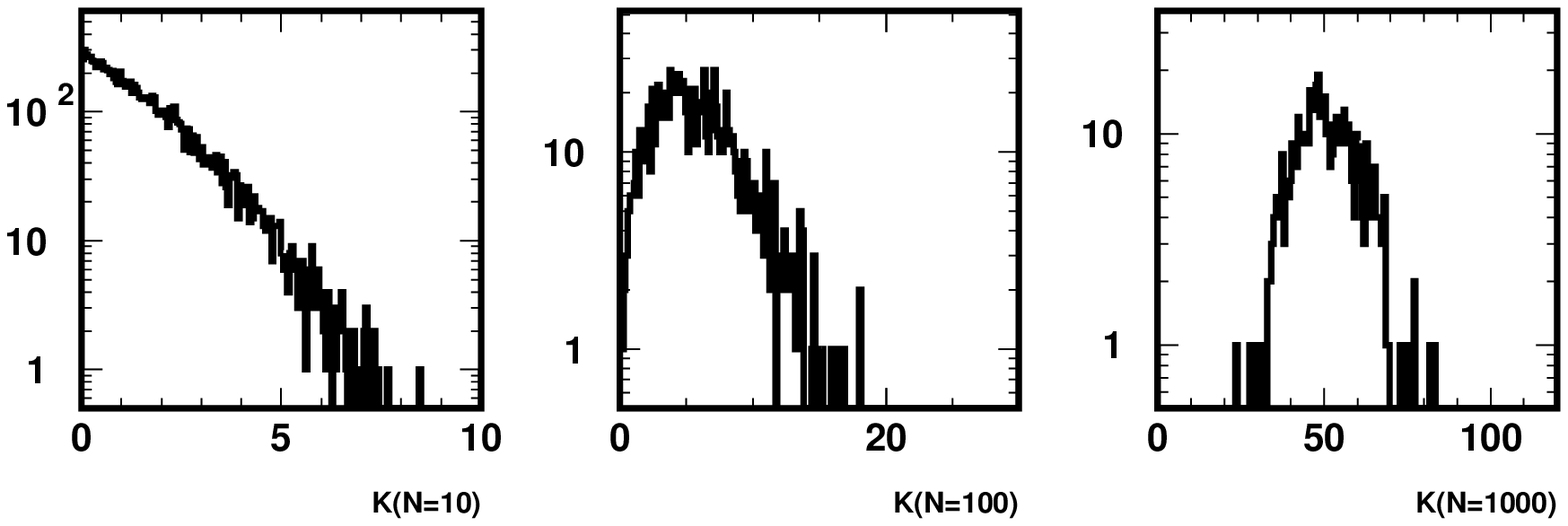}
\caption{
(top row) Distributions in ${K}_1$ for flat PDF: experiments with $N=10$, $N=100$, and $N=1000$, shown with fits to an exponential form.
(bottom row) Distributions in ${K}_1$ for PDF with the form $0.3+1.4X$ with $N=10$, $N=100$, and $N=1000$.
}
\label{fig:null}
\end{figure}
\end{center}
%%%%%%%%%%%%%%%%%%%%%%%%%%%%%%
\end{widetext}

To evaluate rejection power, these distributions may be compared with those obtained for PDF's that are not flat.
The alternative hypotheses used in a study by Aslan and Zech~\cite{Zech} provide a convenient range of function types and allow for a direct comparson with the range of tests reviewed in their work.
In that paper the rejection power of the alternative hypothesis was defined as one minus the probability for an error of the second kind, given a criterion that yields a 5\% significance for the null hypothesis.
Since in this case the null hypothesis gives an exponential distribution with unit decay constant, the 5\% criterion is $K_1>3.0$.
Ensembles of experiments were generated for each of three functions used in Ref.~\cite{Zech}:
\begin{eqnarray}
{\cal A}_1(X)&=& 0.3+1.4X\\
{\cal A}_2(X)&=& 0.7+0.3[n_2e^{-64(X-0.5)^2}]\label{eqn:A2}\\
{\cal A}_3(X)&=& 0.8+0.2[n_3e^{-256(X-0.5)^2}]\label{eqn:A3}
\end{eqnarray}
where the $n_i$ are normalization constants for the associated Gaussians.
All functions are defined in the interval $[0,1]$.
The resulting $K_1$ distributions for ${\cal A}_1$ are shown in Figure~\ref{fig:null}  (bottom row).
The values for rejection power are summarized in Table~\ref{tab:Aflat}.
For comparison, the values for the $\chi^2$ method ($N=100$) given by Ref.~\cite{Zech} are approximately 0.81, 0.85, and 0.81, respectively, so our method is comparable in power, at least in the case of these three functions.

%%%%%%%%%%%%%%%%%%%%%%%%%%%%%%
  \begin{table}[t]
   \begin{center}
    \begin{tabular}{lccc}
     \hline
     \hline 
     Function & \multicolumn{3}{c}{Rejection Power }\\
      & $N=10$ & $N=100$ & $N=1000$
     \\ \hline
     ${\cal A}_1$ (Linear)& 0.117	& 0.824 & 1.00      
          \\
          ${\cal A}_2$ (Wide Gaussian) & 0.152 & 0.910 & 1.00
          \\
          ${\cal A}_3$ (Narrow Gaussian) & 0.102 & 0.672 & 1.00 \\
     \hline
     \hline
    \end{tabular}
    \caption{Rejection power for functions ${\cal A}_1$, ${\cal A}_2$, and ${\cal A}_3$ with a flat null hypothesis.}
    \label{tab:Aflat}
   \end{center}
 \end{table}
%%%%%%%%%%%%%%%%%%%%%%%%%%%%%%

In order to apply this method as a goodness-of-fit test for non-uniform null hypotheses the PDF, $f(X)$, must first be transformed to a ``flat'' variable, $Y$, where the probability distribution is flat.
To form a uniform null hypothesis on a circle one could, for example, construct $Y$ as :
\begin{eqnarray}
Y_i &=& 2\pi{\int_{X_-}^{X_i}f(X)dX}
\label{eqn:flatten}
\end{eqnarray}
where the integer subscript $i$ denotes the $i^{th}$ data point and $X_-$ is the lowest possible value of $X$.
%The $K_1$ distribution of data flattened according to the generator PDF should fit an exponential no differently than that of a flat PDF.
%As numerical approximations in computation may produce distortions, it is important to test each code for this property.

\begin{widetext}
%%%%%%%%%%%%%%%%%%%%%%%%%%%%%%
\begin{center}
\begin{figure}[htb]
\smallskip
\includegraphics[width=135mm]{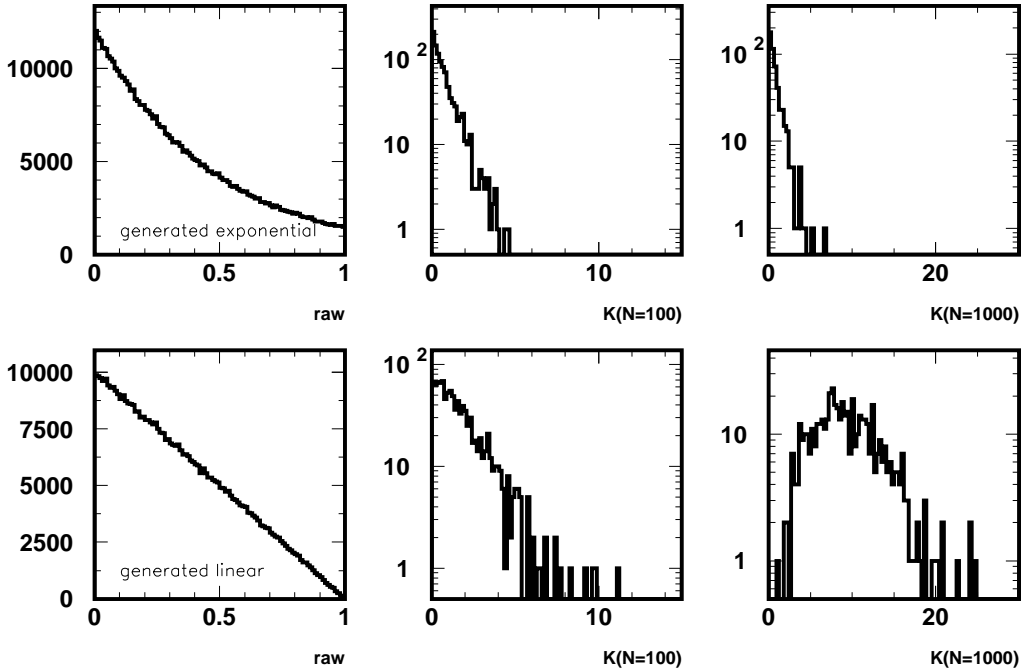}
\caption{Determination of rejection power for a compound hypothesis: ensembles fitted for decay constant of exponentially decaying form.
(top row) PDF matches fit parametrization: (left) Raw distribution, (center, right) distributions in ${K}_1$ of fitted, flattened experiments, $N=100$ and $N=1000$.
(bottom row) PDF inconsistent with parametrization: (left) Raw distribution, (center, right) distributions in ${K}_1$ of fitted, flattened experiments, $N=100$ and $N=1000$.
}
\label{fig:compound}
\end{figure}
\end{center}
%%%%%%%%%%%%%%%%%%%%%%%%%%%%%%
\end{widetext}
\section{Compound hypotheses}
The examples considered thus far have been ones where no parameter fitting has occurred.
While this has been an instructive exercise, it has limited application, as most measurements in particle physics involve the fitting of measured distributions to determine shapes and to derive some physics quantity or conclusion.
We now look at compound hypotheses.

In evaluating rejection of alternative hypotheses via toy MC in the compound case, it is important that the fitting process be integrated into the evaluation procedure.
Consider a data set $\{\phi_i\}$ where the PDF is assumed to be parametrizable as $f(\phi;\alpha)$ and the unbinned likelihood is maximum for $\alpha=\alpha_{max}$.
The data are then flattened assuming the PDF is $f(\phi;\alpha_{max})$, and the associated $K_1$ is evaluated.
The confidence level of this $K_1$ value may then be found by referencing the  ensemble distribution of $K_1$ when the true PDF is  $f(\phi;\alpha_{max})$,
and each experiment of the ensemble is treated as data, fitted and flattened according to the fit.

This procedure was used to evaluate rejection power for pairs of similarly shaped PDF's. 
Here we show one such result, for
the hypothesis $n_4(\alpha)e^{-10X\alpha}$, where $n_4$ is a normalization constant, the measured quantity is $X$, and experiments are fitted for $\alpha$.
The alternative PDF was the linear form $f(X)=2(1-X)$.   
Experiments were generated according to the alternative PDF (A), and each was fitted to the hypothesis.
The mean maximum likelihood value of $\alpha$ was approximately 4.7.
Ensembles (B) were generated according to the hypothesis, with $\alpha=4.7$, and fitted in the same way.
The 5\% confidence criterion on $K_1$ for (B) and acceptance of this criterion for (A) were estimated by counting (Figure~\ref{fig:compound}).  
The rejection powers were found to be 28\% and 99\% for $N=100$ and $N=1000$, respectively.
For comparison we also calculated by the same procedure the rejection of the $\chi^2$ test, using 20 bins in the interval [0,1] and found powers of 13\% and 100\%, respectively.

%%%%%%%%%%%%%%%%%%%%%%%%%%%%%%
\begin{figure}[htb]
\includegraphics[width=0.45\textwidth]{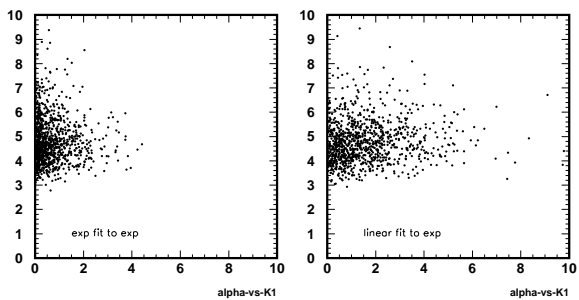}
\caption{
Scatter plots of fitted parameter $\alpha_{max}$  {\it vs.} $K_1$ for ensembles  shown in Figure~\ref{fig:compound} ($N=100$).
}
\label{fig:alphk1}
\end{figure}
%%%%%%%%%%%%%%%%%%%%%%%%%%%%%%

We also examined the two-dimensional distribution of fitted $\alpha_{max}$ values $vs.$ $K_1$.
Any dependence of the test on the fitted rather than underlying parameter value reduces its utility as a goodness-of-fit test;
for example, the maximum likelihood value, ${\cal L}_{max}$, is not usable as a goodness-of-fit statistic because it depends strongly on the fitted parameter value(s) $\alpha_{max}$ -- for a certain class of fitting functions, the correlation is 100\%\cite{likelihood}.
Figure~\ref{fig:alphk1} shows scatter plots of $\alpha_{max}$ and $K_1$, where the data were generated with $N=100$ and the generated and fitted forms are those from the example of Figure~\ref{fig:compound}.
There appears to be no strong dependence.

In any determination of rejection power with a compound hypothesis, it is necessary to determine the distribution of $K_1$ for the correct hypothesis.
It does not appear that there is a simple ansatz as in the case of binned least squares fitting, where the chisquare converges to a chisquare distribution with the number of degrees of freedom reduced by one unit for each linear fitted parameter.
We study this question empirically by generating MC ensembles for a variety of shapes.
Each ensemble was generated according to the fitted functional form with parameter value(s) fixed.
Each experiment was fitted with parameter(s) floating, and the $K_1$  value was obtained from the data flattened according to the best fit.
The distribution of resultant $K_1$  values for each ensemble was fitted for the decay constant, assuming an exponentially decaying form.  
Ensembles with $N=10$, $N=100$, and $N=1000$ were generated.
The results are summarized in Table~\ref{tab:Afitted}.
There are several notable features.
First, while all of the $K_1$ distributions had a decaying form, as one might expect, and a fit that converged, not all yielded good fits; the exponential form is not preserved under compound hypotheses.
Secondly, all inverse decay constants are greater than unity, indicating that the $K_1$ distribution moves toward zero with fitting.
This is not suprising;  fitting identifies for each experiment the shape that is ``closest'' to the data, giving in general a better goodness-of-fit than the generator shape.
Finally, there is no obvious pattern in the value of the decay constant with number of floated parameters.
However, it is seen that for a given PDF and set of fitted parameters, the shape of the $K_1$ distribution shows remarkably little change as $N$ is changed by two orders of magnitude.

\begin{widetext}
%%%%%%%%%%%%%%%%%%%%%%%%%%%%%%
   \begin{center}
  \begin{table}[ht]
    \begin{tabular}{cccccc}
     \hline
     \hline 
     Form & Generated & Fitted &\multicolumn{3}{c}{$K_1$ (Decay Constant)$^{-1}$ ($\chi^2/ndf$)}
\\
   & & & $N=10$ & $N=100$ & $N=1000$
     \\ \hline
         $(1-\alpha)+\alpha (2X)$  & $\alpha=0.7$ & $\alpha$ & -- & -- &  $1.28\pm 0.07$ (70/67)
          \\
 $(1-\alpha)+\alpha [n_2e^{-64(X-0.5)^2}]$ &$\alpha=0.3$ & $\alpha$ & -- & $1.90\pm 0.06$ (230/80) & $1.94\pm 0.09$ (223/65)
\\
$(1-\alpha)+\alpha [n_3e^{-256(X-0.5)^2}]$ &$\alpha=0.2$ & $\alpha$ & --& $1.56\pm 0.05$ (203/82)& $1.56\pm 0.07$ (82/68)
\\
$n_4e^{-10X/\alpha}$ & $\alpha = 1.0$ & $\alpha$ & $1.23\pm 0.01$ (147/133) & $1.28\pm 0.04$ (68/85) & $1.28\pm 0.06$ (75/76)
\\
$n_5e^{-[X-(0.5+\alpha_2)]^2/2(\alpha_1/8)^2}$ & $\alpha_1 = 1.0,$ & $\alpha_1$ & $1.36\pm 0.01$ (176/131) & $1.38\pm 0.05$ (93/85) & $1.50\pm 0.07$ (56/65)
\\
&  $\ \ \ \alpha_2=0$& $\alpha_2$ & $1.22\pm 0.01$ (154/135) & $1.25\pm 0.04$ (122/96) & $1.28\pm 0.06$ (73/72)
\\
& & $\alpha_1,\alpha_2$ & $1.84\pm 0.019$ (148/90) & $2.00\pm 0.065$ (53/59) & $2.13\pm 0.095$ (47/47)
\\
     \hline
     \hline
    \end{tabular}
    \label{tab:Afitted}
    \caption{Inverse decay constants of $K_1$ distribution for several generated forms, flattened after fitting for parameter(s) $\{\alpha_i\}$. 
The $n_i$ are normalization constants, which may depend on the parameters $\alpha_j$.
No entry is made for samples where low statistics resulted in best fits which were at the limits of the parametrization.
}
 \end{table}
   \end{center}
 %%%%%%%%%%%%%%%%%%%%%%%%%%
 \end{widetext}

\section{Extension to multidimensional data: speculation}
Our goal in this investigation has been to arrive at a multidimensional unbinned goodness-of-fit test, one that has rejection power in all dimensions, not just in one-dimensional projections, for multidimensional data.
Many unbinned tests depend on the integrated sum of or spacings between neighboring data points, quantities which are not well-defined when extended to more than one dimension.
Although the $K_1$ statistic does not have this property, it is yet to be determined whether  there exists an extension that is fully multidimensional; for example, in two dimensions, two components each mapped to a circle corresponds to a data space that is the surface of a toroid, for which there is no obvious nontrivial vector sum that maps to the Random Walk.
A fully general extension to multidimensional data will additionally require a flattening algorithm and provisions for data spaces of arbitrary shape.
We will continue to explore the possibilities for extending $K_1$ for use with multidimensional data.

\section{Summary}

We have explored an unbinned goodness-of-fit test for data in one dimension that is based on the mapping of flattened distributions to a two-dimensional random walk.
This method is truly binning-free and scale-independent, and the ensemble distribution for the null hypothesis is well-defined.
For a compound hypothesis we specify a procedure to determine the
ensemble distribution of the test statistic via Monte Carlo so that rejection power may be readily determined.
The distribution is found for several different parametrized forms and
shown to be largely independent of statistics.
We examine several samples for dependence between the test statistic and fitted parameter values, and find no evidence of any.
The rejection power for alternate hypotheses is demonstrated for a few examples and is found to be comparable to that of the chisquare method.

\smallskip
\begin{acknowledgments}

The author would like to thank R. Cousins, G. Zech, and B. Yabsley for useful discussions and suggestions, and the organizers of PHYSTAT 2003 for a stimulating and interesting conference.  
This work is supported by Department of Energy grant  {\#}DE-FG02-84ER40153.  
\end{acknowledgments}

\bigskip

\end{document}